\documentclass[letterpaper,final,12pt]{article}
\usepackage{graphicx}
\usepackage{graphics}

\begin{document}

\author{K. Yu. Bliokh\footnote{Radio Astronomy Institute, National Academy of Scienses of Ukraine, 61002 Kharkov, Ukraine, e-mail: k\_bliokh@mail.ru} and Yu. P. Bliokh\footnote{Departmant of Physics, Technion, 32000 Haifa, Israel, e-mail: bliokh@physics.technion.ac.il}}
\date{}
\title{What are the left-handed media and what is interesting about them?\footnote{Published as Methodological Notes in {\it Physics -- Uspekhi} {\bf 47} (4) 393 (2004).}}

\maketitle
\begin{abstract}
We review the intensively discussed ideas about 
wave propagation and refraction in media where both electric permittivity and magnetic permeability are negative. The criticism against negative refraction as violating the causality principle is considered. Starting from the initial wave 
equations, refraction of beams at the boundary of a left-handed medium is analyzed. The physics of a perfect lens 
formed by a flat layer of a left-handed material is considered. 
\end{abstract}

\tableofcontents

\section{Introduction}

\begin{figure}[tbh]\centering
\scalebox{0.5}{\includegraphics{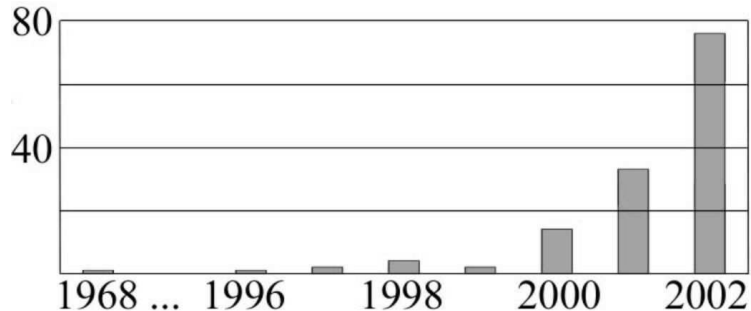}}
\caption{ The number of publications on negative-index media (NIM) 
(from Ref.~\cite{Pendry1})}
\label{Fig1}
\end{figure}

The term ``left-handed media'' was first introduced by 
V. G. Veselago in 1967 \cite{Veselago1} for media with a negative refractive 
index\footnote{The concept of a negative refractive index was also first introduced by 
Veselago \cite{Veselago2}}. The increase in the number of publications on this 
subject is clearly illustrated by the table (Fig.~\ref{Fig1}) published by 
J. B. Pendry in April 2003 in the introduction to a special issue 
of Optical Express \cite{Pendry1}. A flood of interest in this problem at 
the end of the 1990s was initiated by the creation of composite 
media with negative refraction and their first experimental 
studies. What is so interesting about these media that has 
made many physicists abandon their current activities and 
turn to investigating them? The present paper is an attempt to 
answer this question. 

\section{Negative-index media}

The sign of the refractive index of a medium depends on 
whether the phase and group velocities of a wave are parallel 
or antiparallel in this medium. In the first case, the group 
velocity is considered positive and in the second case it is 
considered negative. We believe that it is worth quoting 
L. I. Mandel'shtam's paper of 1945 \cite{Mandelshtam} in this connection: 

``Perhaps it should be stressed that the sign of the group 
velocity has a considerable effect on some phenomena that 
are usually discussed without even mentioning the group 
velocity. 

I mean, for instance, reflection and refraction of a plane 
wave at the plane boundary between two nonabsorbing 
media. 

In deriving the corresponding equations -- for the 
direction of the refracted beam and for the amplitudes of 
the reflected and refracted waves -- one makes an 
assumption, usually not mentioned explicitly, that the 
phase velocity of the refracted wave forms an {\it acute} angle 
with the normal to the boundary directed towards the 
``second'' medium. 

At the same time, according to the physical essence of the 
problem, this assumption should relate to the group velocity 
(to the velocity of energy propagation). The resulting 
equations are correct only because in real situations, as 
mentioned above, we deal with positive group velocities.

If the group velocity is negative, the requirement that the 
energy propagates from the boundary is equivalent to the 
requirement that the phase on the boundary is increasing. In 
this case, the refracted beam is directed not in a usual way but 
symmetrically with respect to the normal to the boundary.'' 

In the same paper, Mandel'shtam notes that a spatially 
periodic medium (in his analysis, a crystal lattice) provides an 
example of a medium where the refractive index can be 
negative within a certain frequency range. Composite left-handed media, which were developed at the end of the nineties \cite{Pendry2, Pendry3}
and caused an explosion of interest in this problem, are 
also spatially periodic. Such media have a negative refractive 
index for microwaves with frequencies of the order of 10~GHz. 

Periodic waveguide systems, or slow wave structures, have 
been well known in microwave electronics for a long time. 
Waves with negative refractive index are also well known: 
these are backward waves, or negative-dispersion waves. So 
what is the difference between the ``old'' slow wave structures 
and the new ones, the left-handed media? A principal 
difference is that the traditional ``old'' slow wave structures 
are {\it one-dimensional}, while the left-handed media are {\it multi-dimensional} (two- or three-dimensional). Refraction of a wave 
at the interface of two media is a multi-dimensional effect and 
is therefore absent in slow wave structures. 

Investigation of multi-dimensional periodic structures 
was started long ago (see Refs.~\cite{Silin1,Silin2,Silin3} and the references 
therein) but the authors of Refs.~\cite{Pendry2, Pendry3} do not cite any previous 
works. This is possibly because they are unaware of papers 
published in Russian; another reason may be that their 
approach to the problem is totally new. The authors of 
theoretical paper~\cite{Pendry2} consider the possibility of creating a 
periodic structure with a surface where low-frequency weakly 
decaying waves could exist, similarly to plasmons -- collective waves in metals. This is possible in a medium with 
negative electric permittivity (dielectric constant). The 
authors of Ref.~\cite{Pendry2} have shown that a periodic grating of 
thin wires has low losses and negative electric permittivity 
$\varepsilon<0$ for electromagnetic waves with frequencies of 1--10~GHz. Later, it was shown~\cite{Pendry3}  that a periodic grating of 
ring cavities has negative magnetic permeability $\mu<0$ in the 
same range. By combining both gratings, a left-handed 
negative-index medium was formed~\cite{Smith1,Shelby1}. 

It should be said to the credit of Western scientists that 
they managed to make their research a tremendous success 
due to publications in respected physics journals as well as in 
popular publications, including newspapers\footnote{A large and constantly updated collection of references can be found at http://physics.ucsd.edu/$\sim$drs/left\_home.htm.}.
But we now put the questions of prestige aside and turn to physics. 

As another example~\cite{Notomi} of a left-handed medium, we 
mention photonic crystals, which are media with a spatially 
periodic refractive index $n({\bf r})$. 

Thus, left-handed media are two- or three-dimensional 
periodic structures. In such anisotropic media, the angle 
between the phase and group velocities of a wave can be 
different. The only exceptions are spatially periodic media 
that are {\it isotropic}\footnote{While we were preparing this material for publication, a paper appeared~\cite{Zhang} where refraction of light at the interface of two anisotropic media with different directions of optic axes was erroneously interpreted as a 
manifestation of the negative refractive index (for more details, see 
Ref.~\cite{Bliokh}) } in the {\it long-wave} approximation, when the 
wavelength $\lambda$ is much larger than the period $d$, $\lambda\gg d$. In 
this case, a medium can be characterized by effective electric 
permittivity $\varepsilon$ and effective magnetic permeability $\mu$, and 
waves can be classified as forward and backward. It was 
shown in Ref.~\cite{Notomi} that in photonic crystals, where 
modulation of parameters is rather strong, anisotropy in 
some frequency ranges near bandgaps can be negligibly 
small. 

\section{Wave propagation and refraction 
in left-handed media}

We consider a plane electromagnetic wave propagating in a 
medium with a scalar electric permittivity $\varepsilon$ and a magnetic 
permeability $\mu$. If $\varepsilon>0$ and $\mu>0$, the electric field $\vec E$, 
magnetic field $\vec H$, and wavevector $\vec k$ form a {\it right-handed} 
vector triplet; if $\varepsilon<0$ and $\mu<0$, they form a {\it left-handed} 
triplet. This is the origin of classifying media into left-handed 
and right-handed~\cite{Veselago1}. The energy flux of a wave is given by the 
Poynting vector
\[\vec S=\frac{c}{4\pi}\left[\vec E \vec H\right]~,\]
which always forms a right-handed triplet with the vectors $\vec E$ 
and $\vec H$. Therefore, the group and phase velocities are parallel 
in right-handed media (positive group velocity) and are anti-parallel in left-handed media (negative group velocity).

\begin{figure}[tbh]\centering
\scalebox{0.5}{\includegraphics{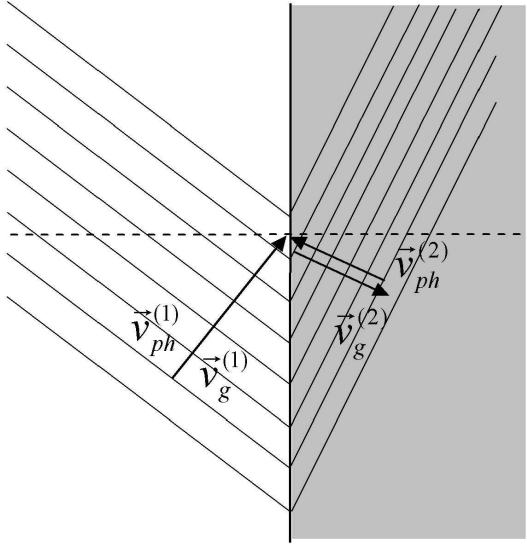}}
\caption{Refraction of a plane wave. }
\label{Fig2}
\end{figure}

Refraction of a plane wave at the interface of a left-handed medium and a right-handed one looks quite unusual. 
For definiteness, we consider a wave propagating from a 
usual right-handed medium (which is assumed to be the 
vacuum in what follows) to the plane boundary of a left-handed medium. The group velocity of the refracted wave is 
directed from the boundary and its phase velocity is directed 
towards the boundary. This means that the phase velocities of 
both waves are directed towards the boundary, and hence the 
phases of both waves increase on the boundary. This is 
possible only if the propagation directions of both waves 
(group velocities) are {\it on the same side of the normal} to the 
boundary (Fig.~\ref{Fig2}). In other words, the refractive index $n$ 
entering Snell's law for the left-handed medium is {\it negative}, $n=-\sqrt{\varepsilon\mu}$.

\begin{figure}[tbh]\centering
\scalebox{0.5}{\includegraphics{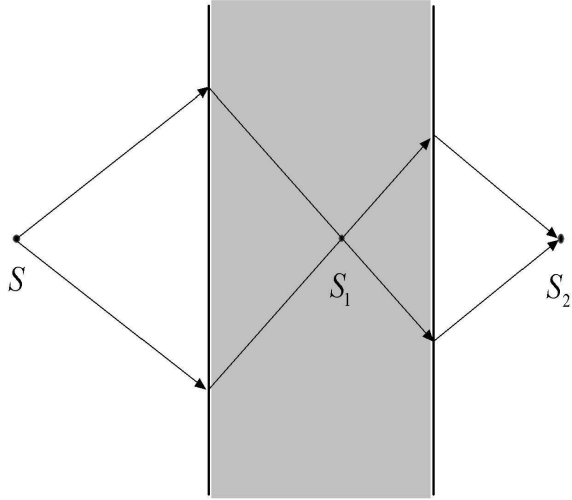}}
\caption{A plane-parallel plate of a left-handed material acts as a 
collecting lens.}
\label{Fig3}
\end{figure}

\begin{figure}[tbh]\centering
\scalebox{0.5}{\includegraphics{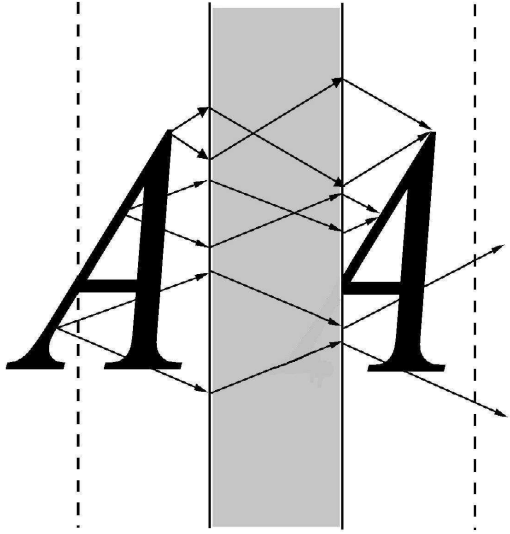}}
\caption{Athree-dimensional image obtained by means of a plane-parallel 
plate made of a left-handed material.}
\label{Fig4}
\end{figure}

The fact that the incident and refracted waves are on the 
same side of the normal to the boundary enables one to 
manufacture quite unusual optic elements of left-handed 
media. For instance, a plane-parallel plate made of a left- 
handed material works as a collecting lens, as can be readily 
seen from the ray diagram in Fig.~\ref{Fig3}. Such a lens has a 
remarkable feature: it has no focal plane. It therefore forms 
a {\it three-dimensional} image of an object, which makes it similar 
to a mirror. But in contrast to a mirror, it forms a real image, 
and this feature opens new possibilities for three-dimensional 
photography. Of course, this flat lens also has a certain 
drawback: for an object to be imaged, it must be placed 
sufficiently close to the surface of the lens. For instance, for an 
object placed near a lens made of a ``perfect'' left-handed 
material ($\varepsilon=\mu=-1$), only those points whose distance 
from the lens surface does not exceed the thickness of the 
lens have real images (Fig.~\ref{Fig4}). 

It is no accident that we have used the term ``perfect left-handed medium'' for the medium with $\varepsilon=\mu=-1$. Indeed, 
such a medium has some additional interesting features. First, 
a perfect left-handed medium has zero reflection coefficient: 
all the energy of the incident wave is passed to the refracted 
wave. Second, a plane-parallel plate of a perfect left-handed 
material forms a perfect image because the phase incursion 
along any trajectory between the object and the image equals 
zero. This can be easily understood by noting that for any 
beam traveling from the object to the image, half of the path is 
in the ordinary medium and the other half is in the perfect left-handed medium. Because phase velocities in the two media 
have equal values but opposite directions, phase delays along 
the two parts of any trajectory {\it exactly} compensate each other. 

But these are not all the wonderful properties of a perfect 
left-handed medium. In 2000, Pendry published a paper 
``Negative refraction makes a perfect lens'' \cite{Pendry4}. Here, ``a perfect lens'' means a lens whose resolving power {\it exceeds the 
limit set by the wave nature of light}. In order to understand the 
author's reasoning, we consider a monochromatic light 
source placed in the plane parallel to a plate of a perfect left-handed material. Let the distribution of the field (magnetic 
field, for instance) in the source plane be given by a function 
$H_0(x)$. This function can be represented as a spatial Fourier 
transform, 
\begin{equation} \label{1}
H_0(x)=\intop_{-\infty}^{+\infty}H_k(k_\perp)\exp(ik_\perp x)dk_\perp~.
\end{equation}

The field of the wave propagating along the $z$ axis from the 
source to the plate can be represented as
\begin{equation}\label{2}
    H(x,z)=\intop_{-\infty}^{+\infty}H_k(k_\perp)\exp(ik_\perp
    x)\exp\left(i\sqrt{\omega^2/c^2-k_\perp^2}z\right)dk_\perp~.
\end{equation}

It can be seen from representation (\ref{2}) that the Fourier 
components with $k_\perp>\omega/c$ decay exponentially with the 
distance from the source. These nonpropagating waves are 
usually neglected. Omitting harmonics with $k_\perp>\omega/c$ means 
losing the information about the details of the image that are 
smaller than the wavelength $\lambda=2\pi c/\omega$. 

In Ref.~\cite{Pendry4}, it is noted that the exponential decay of nonpropagating waves affects only their amplitudes but not their 
phases. Therefore, one can completely restore the information about the spatial structure of the source by amplifying 
these waves. It is shown in Ref.~\cite{Pendry4} that a plate of a perfect 
left-handed material can serve as this ``perfect'' amplifier, 
increasing the amplitudes of nonpropagating waves by 
exactly the same value as that of their decay. 

These unusual (and quite tempting from the practical 
standpoint) properties of left-handed materials have caused 
an outburst of publications during the recent years. However, 
not all the participants in the discussion in the literature agree 
with the optimistic conclusions that we have briefly reviewed 
in this section.

\section{Negative refraction and the causality principle}

\begin{figure}[tbh]\centering
\scalebox{0.5}{\includegraphics{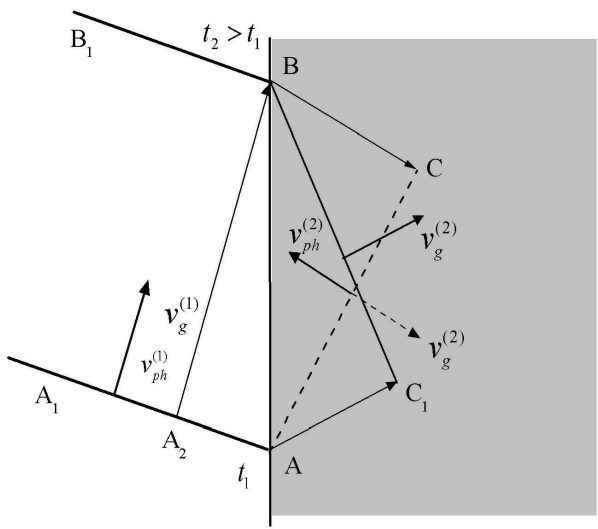}}
\caption{Negative refraction and the causality principle.}
\label{Fig5}
\end{figure}

In 2002, {\it Physical Review Letters} published a paper entitled 
``Wave refraction in negative-index media: always positive 
and very inhomogeneous'' \cite{Valanju}. The authors claim that 
negative refraction contradicts the causality principle and 
therefore does not exist. The authors' reasoning can be most 
easily understood with the help of Fig.~\ref{Fig5}.

Let a {\it wave packet} be incident on an interface of the 
vacuum (left) and a left-handed isotropic medium (right). 
The group velocity $\vec v_g^{(1)}$ and the phase velocity $\vec v_{ph}^{(1)}$ of the wave
packet in the vacuum coincide. Let the wave packet front 
coincide with the line $AA_1$ at time $t_1$ and with the line $BB_1$ at 
time $t_2>t_1$. In the left-handed medium, the phase velocity of the refracted wave $\vec v_{ph}^{(2)}$ is directed towards the boundary. 
According to the authors' reasoning, if the group velocity $\vec v_g^{(2)}$ 
of the refracted wave were directed oppositely to the phase 
velocity, then the front of the refracted wave packet would be 
oriented along the line $AC$ (dashed line in Fig.~\ref{Fig5}), orthogonally to the group velocity direction. Because the time required 
for the front displacement along the path $AA_1\rightarrow BB_1\rightarrow AC$
is the same for all points of the front, it follows from Fig.~\ref{Fig5} 
that the signal should travel along the path $A_2BC$ with infinite 
velocity. This violates the causality principle. The authors 
conclude that the front of the refracted wave should coincide 
with the bold line $BC_1$ and the group velocity should be 
orthogonal to it. They suggest that one should distinguish 
between phase and group refractive indices. The first 
describes the change in the direction of the phase velocity, 
and it is indeed negative for left-handed media. But the group 
refractive index describes the direction of energy propagation, and it is positive. Because the beam diagram used in 
geometric optics (like the one shown in Fig.~\ref{Fig3}) should indicate 
the direction of energy propagation and not phase propagation, there can be no focusing properties of the plate, to say 
nothing of forming a ``perfect'' image.

Further, a large difference between the group and phase 
velocities of the refracted wave means strong dispersion; 
therefore, a wave packet ``spreads'' after passing the interface 
of the two media, its amplitude decreases very fast, and decay 
of the incident wave occurs instead of focusing. 

The standpoint of the authors of Ref.~\cite{Valanju} is described in 
full detail, with pictures and animations, at http://www.utexas.edu/research/cemd/nim/. 

\section{Wave and geometric optics 
of a left-handed me\-di\-um}

Two flaws can be seen in the above proof that the ``actual'' (i.e., 
group) refraction cannot be negative. First, the authors claim 
that group and phase velocities can be noncollinear in an 
isotropic medium. Second, to investigate the focusing properties of an optical system (in our case, a plane-parallel plate) 
one does not need to consider a multi-frequency wave packet, 
hence, dispersion of the medium is not relevant. 

In order to find the source of error in Ref.~\cite{Valanju}, we consider 
refraction of a narrow monochromatic light beam in the 
framework of the wave equation. First of all, this approach 
enables us to do without the concept of group velocity direction as the direction of energy propagation. (Indeed, 
the beam direction is evidently the same as the direction of 
energy propagation.) Second, using the wave equation 
approach allows finding whether the geometric optics 
approximation is applicable to left-handed media.

Let the incident wave be the exact solution of the two-dimensional Maxwell equations in the vacuum. It can be 
represented as the Fourier integral
\begin{eqnarray}
E_{x}={1\over\pi}\int_{-\infty}^{\infty}d k_\perp{{\sin (k_\perp
a)}\over k_\perp} {\sqrt{ k_0^2- k_\perp^2}\over k_0}
\exp{\left(i k_\perp x+i\sqrt{k_0^2- k_\perp^2} z\right)}~,\nonumber\\
E_{z}=-{1\over\pi}\int_{-\infty}^{\infty}d k_\perp{{\sin (k_\perp
a)}\over k_0}
\exp{\left(i k_\perp x+i\sqrt{k_0^2- k_\perp^2} z\right)}~,\nonumber\\
H_{y}={1\over\pi}\int_{-\infty}^{\infty}d k_\perp{{\sin (k_\perp a)}\over k_\perp}
\exp{\left(i k_\perp x+i\sqrt{k_0^2- k_\perp^2} z\right)}~.
\label{eq1}
\end{eqnarray}
Here, $k_0=\omega/c$. We consider the two-dimensional case for 
simplicity. Electric and magnetic fields in (\ref{eq1}) form a 
monochromatic wave (we have omitted the factor 
$e^{-i\omega t}$ in all terms). The source of this wave is placed in 
the plane $z=0$ and has the following distribution of magnetic 
field\footnote{This boundary condition was used by Kirchhoff in solving the problem 
of light diffraction by a slit.}:
\begin{eqnarray}
H_0(x)\equiv H_y(x,0)=1\,\,\, {\rm at}\,\,\,\mid x\mid<a\nonumber\\
H_0(x)=0\,\,\, {\rm at}\,\,\,\mid x\mid>a.
\end{eqnarray}
Here, $a$ denotes the size of the source in the $x$ direction. Such a 
wave forms a beam propagating along the $z$ axis. The angular 
divergence of the beam, $\alpha\sim (k_0a)^{-1}$, is small if the wavelength is small compared to the source size, $k_0 a\gg 1$. In what 
follows, we assume this condition to be satisfied. 

The field (\ref{eq1}) at an arbitrary point $(x,\,z)$ consists of 
propagating $(\mid k_\perp\mid<k_0)$ and decaying $(\mid k_\perp\mid>k_0)$ plane 
waves. At a large distance from the source, decaying waves 
can be neglected (we discuss below whether they can be 
amplified in a left-handed medium), and the integrals in (\ref{eq1}) 
must be calculated for $-k_0< k_\perp<k_0$. To simplify the 
notation, we omit the integration limits in what follows. 

\begin{figure}[tbh]\centering
\scalebox{0.5}{\includegraphics{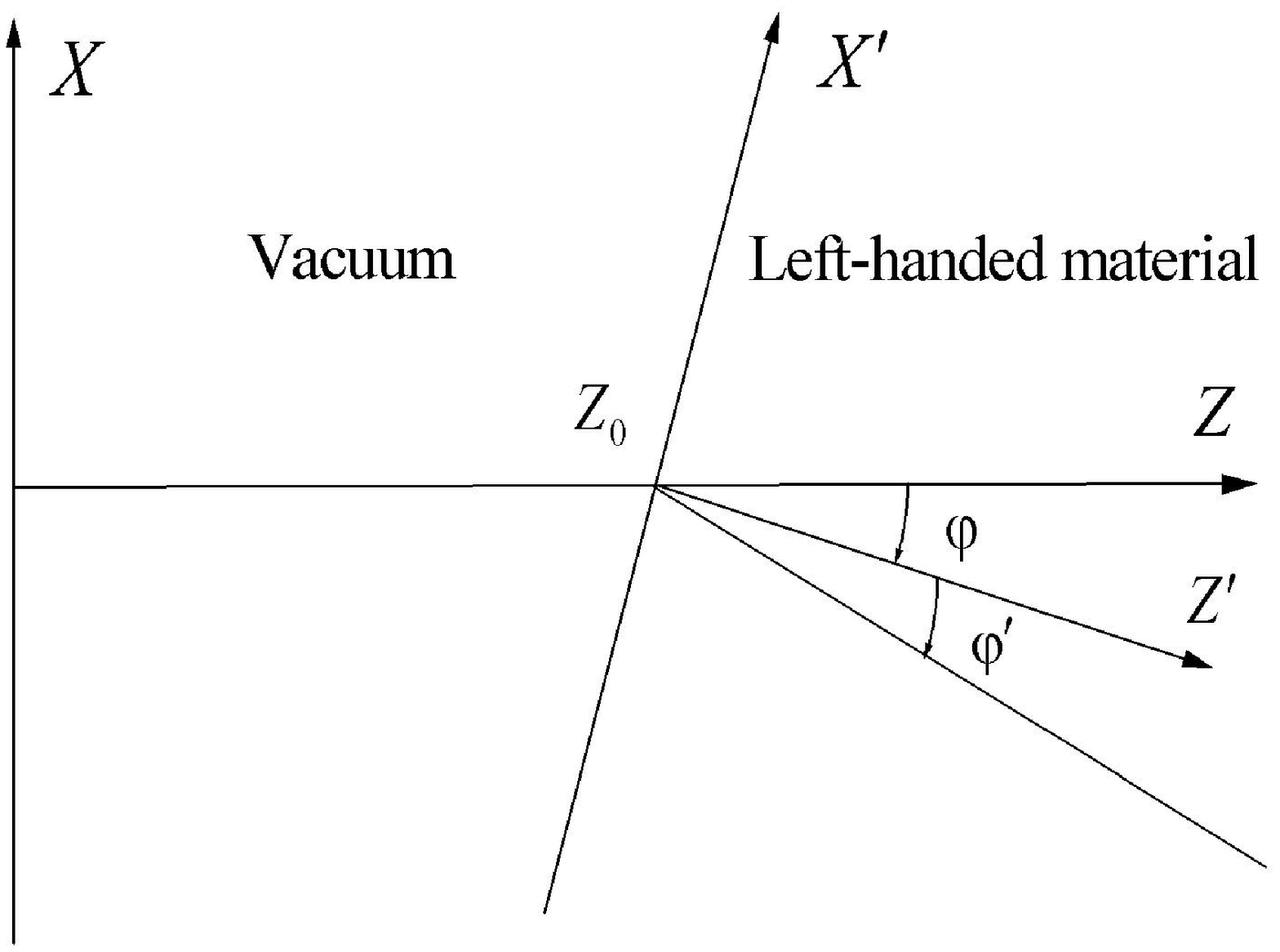}}
\caption{}
\label{Fig6}
\end{figure}

Let the plane boundary between the vacuum and the 
medium intersect the $z$ axis at a point $z=z_0$ and the normal to 
the surface form the angle $\varphi$ with the $z$ axis (Fig.~\ref{Fig6}). We 
introduce the new coordinates $x^\prime$, $z^\prime$ with the origin at the 
point $(0, z_0)$. The $x^\prime$ axis is directed along the boundary and 
the $z^\prime$ axis along the inner normal. At the boundary $z^\prime=0$, 
the magnetic field of the incident wave can be written as 
\begin{eqnarray}
H_{y}^{(in)}={1\over\pi}\int d k_\perp^\prime{\sqrt{k_0^2-
k_\perp^2(k_\bot^\prime)}\over \sqrt{k_0^2-{k_\perp^\prime} ^2}}{{\sin
\left[k_\bot(k_\bot^\prime)a\right]} \over k_\bot(k_\bot^\prime)} \exp{\left[i
k_\bot^\prime x^\prime+i z_0\sqrt{k_0^2- {k_\bot}^2(k_\bot^\prime)}\right]}~, \label{eq2}
\end{eqnarray}
где 
\[k_\bot(k_\bot^\prime)=k_\bot^\prime\cos\varphi-\sin\varphi
\sqrt{k_0^2-{k_\bot^\prime}^2}.\]
Fourier components of the electric field can be expressed in 
terms of the magnetic field components, and we do not 
therefore consider them here explicitly. 

Using the boundary conditions 
\begin{eqnarray}\label{bound}
E_t^{(in)}(k^\prime_\bot)+E_t^{(ref)}(k^\prime_\bot)=E_t^{(tr)}(k^\prime_\bot)~,\nonumber\\
H_t^{(in)}(k^\prime_\bot)+H_t^{(ref)}(k^\prime_\bot)=H_t^{(tr)}(k^\prime_\bot)
\end{eqnarray}
for the Fourier components of the tangential and normal 
parts of the fields, we can calculate the amplitudes of the 
Fourier components of the reflected (labeled by ``ref'') and the 
refracted (labeled by ``tr'') waves on the surface and, hence, at 
any space point $x^\prime, z^\prime$. This procedure is standard, and the 
only peculiar feature in the case of a left-handed medium is 
the unusual sign of the normal part
\[k^\prime_\parallel=\pm\sqrt{k_0^2\varepsilon\mu-{k^\prime_\bot}^2}\]
of the refracted wave wavevector. As we have mentioned 
above, the sign is chosen such that the energy flux in the 
refracted wave is directed from the boundary. In a left-handed 
medium, the wavevector, the magnetic field, and the electric 
field form a left-handed triplet, and hence $k^\prime_\parallel<0$.

Taking into account that at an arbitrary point of the 
medium $x^\prime, z^\prime$, each Fourier component differs from its 
value at the boundary by the factor
\[\exp\left(i\,\rm{sgn}\,\varepsilon\sqrt{k_0^2\varepsilon\mu-
{k^\prime_\bot}^2}z^\prime\right),\]
we easily express the magnetic field of the refracted wave as
\begin{eqnarray}
H_t^{(tr)}={1\over\pi}\int dk^\prime_\bot f(k^\prime_\bot){{\sin
\left[k_\bot(k^\prime_\bot)a\right]}\over k_\bot(k_\bot^\prime)}
\exp{\left[i\Phi(k^\prime_\bot)\right]}~,
 \label{eq3}
\end{eqnarray}
where
\begin{eqnarray*}
 f(k^\prime_\bot)=
{{2|\varepsilon|\left(k^\prime_\bot\sin\varphi+\sqrt{k_0^2-
{k^\prime_\bot}^2}\cos\varphi\right)} \over
{\sqrt{k_0^2\varepsilon\mu-{k^\prime_\bot}^2}+|\varepsilon|
\sqrt{k_0^2-{k^\prime_\bot}^2}}},
\end{eqnarray*}

\begin{eqnarray*}
\Phi(k^\prime_\bot)=z_0\left(k^\prime_\bot\sin\varphi+\sqrt{k_0^2-
{k^\prime_\bot}^2}\cos\varphi\right)+\\
k^\prime_\bot x^\prime+{\rm sgn}\,\varepsilon
\sqrt{k_0^2\varepsilon\mu-{k^\prime_\bot}^2} z^\prime~.
\end{eqnarray*}

Expression (\ref{eq3}) is valid for both right-handed and left-handed media. The only difference is the factor $\rm{sgn}\,\varepsilon$ in the 
phase $\Phi(k^\prime_\bot)$.

We now analyze Eq~(\ref{eq3}). In the simplest case of a perfect 
left-handed medium ($\varepsilon=\mu=-1$) with $\varphi=0$,
\[H_t^{(tr)}(x,z^\prime)=H_y(x,z_0-z^\prime).\]
This relation implies that after crossing the boundary, a 
divergent beam becomes convergent and restores the image 
of the object in the plane $z=2z_0$. In the general case, the 
amplitude of the magnetic field is maximal when the 
stationary point $k_s$, which is determined by the condition 
$d\Phi/dk^\prime_\bot|_{k^\prime_\bot=k_s}=0$, coincides with the point $k_\bot(k_\bot^\prime)=0$ where the function $\sin (k_\bot a)/k_\bot$ is maximal:
\[k_\bot(k_s)=0.\]
This condition is satisfied on the line
\[{x^\prime\over z^\prime}\equiv\tan\varphi^\prime={\rm{sgn}\,\varepsilon
\sin\varphi\over\sqrt{\varepsilon\mu-\sin^2\varphi}},\]
which is natural to interpret as the central line of the refracted 
beam. The amplitude of the wave and, hence, the energy flux 
are maximal along this direction; therefore, the group velocity 
is directed along it. It is easy to see that the angles $\varphi$ and $\varphi^\prime$ are 
related through Snell's law,
\begin{equation}
\frac{\sin\varphi}{\sin\varphi^\prime}=\rm{sgn}\,\varepsilon\, \sqrt{\varepsilon\mu}~,
\label{eq5.5}
\end{equation}
where the sign of $\varepsilon$ indicates whether the incident and 
refracted beams are on the same side of the normal to the 
boundary. We also note that
\[\tan\varphi^\prime={k^\prime_\bot\over k^\prime_\parallel}=
{\rm{sgn}\,\varepsilon\,k^\prime_\bot\over\sqrt{\varepsilon\mu k_0^2-{k^\prime_\bot}^2}},\]
and therefore the refraction law is the same for the phase and 
group velocities.

Thus, by {\it exactly} solving the Maxwell equations, we see 
that there is no difference between phase and group refractive 
indices and that the refractive index is negative for a left-handed medium.

\begin{figure}[tbh]\centering
\scalebox{0.5}{\includegraphics{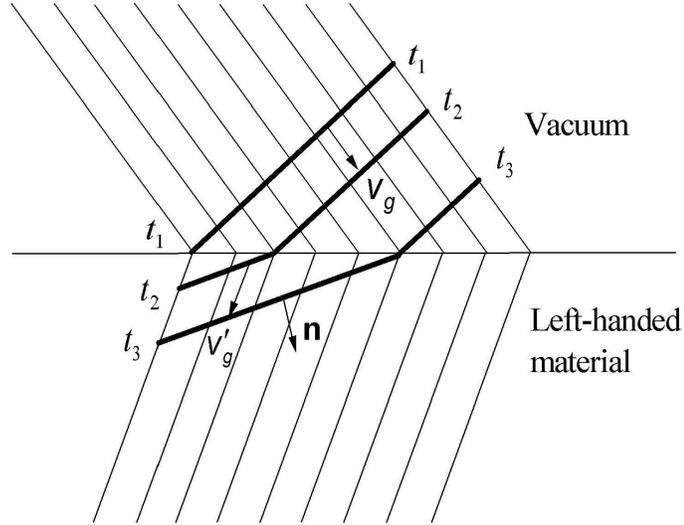}}
\caption{Transformation of a wave packet at the boundary of a left-handed medium.}
\label{Fig7}
\end{figure}

We next consider the contradiction between the exact 
solution of the Maxwell equations and the statement in 
Ref.~\cite{Valanju} that ``causality and finite signal speed preclude 
negative refraction for any waves incident on any material, 
including NIM'' (negative-index medium). We return to the 
scheme of wave packet refraction shown in Fig.~\ref{Fig6}; this time,
however, negative refraction is assumed. Figure~\ref{Fig7} shows the positions of the wave packet at three consecutive time instants, $t_1<t_2<t_3$, which allows reconstructing the shape of the packet in the medium. Indeed, the front of the refracted 
packet coincides with the line $BC_1$ shown in Fig.~\ref{Fig5}, but the 
statement in Ref.~\cite{Valanju} that the normal to the front gives the 
group velocity direction is erroneous. Hence, the conclusion 
that the phase and group velocities in a left-handed medium 
are noncollinear is also erroneous. Actually, when the packet 
crosses the boundary of a left-handed medium, its shape 
changes: an initially ``straight'' packet becomes a ``slanting'' 
one. (In Ref.~\cite{Pendry5}, such behavior was called ``crab-like 
motion''.) Certainly, there is no causality violation here. 

We had two reasons to choose the test solution of the 
Maxwell equations in the form of a narrow wave beam in this 
section. First, it allowed us to resolve the paradox in Ref.~\cite{Valanju}
without using the notion of group velocity; second, this way 
we could formally demonstrate that geometric optics (the 
concept of beams) is applicable to left-handed media. 
Actually, we never used the inequality $k_0 a\gg 1$; therefore, 
the solutions we obtained are also valid for describing other 
situations.

We finally note that after paper \cite{Valanju} had been published, a 
number of papers appeared where the authors, by means of 
analytic \cite{Pendry5,Pacheco} or numerical \cite{Foteinpoulou,Loshiapo} methods, came to the same conclusion as the one we make in this section. However, 
in all those papers, {\it nonmonochromatic} waves were considered. 
In our opinion, this is not necessary for solving the present 
problem and can make the wrong impression that the 
methods of traditional optics, where nonmonochromaticity 
is considered only as a source of chromatic aberrations, 
cannot be applied to left-handed media. 

\section{A perfect lens}

In analyzing equation (\ref{eq3}), we have already considered the 
imaging of a source inside a left-handed medium in the 
simplest case of $\varepsilon=\mu=-1$ and a flat source placed parallel 
to the boundary of the medium. The same relation (\ref{eq3}) can be 
used in a more general case where the source has arbitrary 
orientation relative to the boundary. We express $k^\prime_\bot$ through 
$k_\bot$ and write Eq.(\ref{eq3}) in the form
\begin{equation}\label{eq4}
H_t^{(tr)}={1\over\pi}\int_{-\infty}^{\infty} dk_\bot {{\sin (k_\bot a)}\over k_\bot}
\exp{\left[i\Phi(k_\bot)\right]}~,
\end{equation}
where
\[\Phi(k_\bot)=k_\bot(x^\prime\cos\varphi+z^\prime\sin\varphi)+ \sqrt{k_0^2-k_\bot^2}
(z_0+x^\prime\sin\varphi-z^\prime\cos\varphi).\]
The field given by (\ref{eq4}) has the same form as the source field $H_0(x)$ on the line $z_0=z^\prime\cos\varphi+x^\prime\sin\varphi$:
\begin{eqnarray}\label{eq5}
H_t^{(tr)}\left(x^\prime,z^\prime=z_0/\cos\varphi+x^\prime \tan\varphi\right)=
H_0(z_0\tan\varphi+x^\prime/\cos\varphi)~.
\end{eqnarray}

Relation (\ref{eq5}) means that the source and its image are 
placed symmetrically with respect to the boundary. Similarly, one can obtain relations in a more general case of a 
medium with arbitrary $\varepsilon$ and $\mu$. Without presenting these 
relations, we only mention that the source field can be {\it exactly} 
restored only in the case of a perfect left-handed medium. In 
all other cases, the image has imperfections. Monochromatic 
aberrations can be avoided if the lens is made of an 
anisotropic left-handed material with specially chosen dispersion dependence \cite{Silin4}, but we do not consider anisotropic 
media in the framework of this paper. The properties of 
anisotropic left-handed media are considered in detail in 
Ref.~\cite{Silin3}. 

We recall that the accuracy of restoring the image inside a 
plane layer of a perfect left-handed medium is limited by the 
wavelength, as in usual optical systems. But this is true only in 
the case where the left-handed medium does not amplify 
nonpropagating waves with $k_\perp>k_0$. At first sight, the 
statement in Ref.~\cite{Pendry4} that a layer of a left-handed medium 
amplifies nonpropagating waves and hence enables one to 
exceed the wave limit for the accuracy of image restoration 
looks erroneous because a left-handed medium is a passive 
medium. But we refrain from hasty conclusions and try to 
scrutinize the problem. Because the problem is linear, we can 
restrict consideration to a single Fourier component of the 
field, $\sim\exp{(ik_\perp x)}$. 

We start again from the problem of a plane wave incident 
on an interface of a right-handed medium and a left-handed 
one. Let the magnetic field $H_y$ of the incident wave be 
$H^{(in0)}\exp{(ik_zz)}$, the plane $z=z_l$ being the interface of the media.

The magnetic fields of the transmitted and reflected waves 
occurring at the interface are represented, respectively, as 
\[H^{(tr1)}\exp{(iq_zz)}\,\,\,(z>z_l),\,\,\,\,\,H^{(ref1)}\exp{(-ik_zz)}\,\,\,(z<z_l),\]
where
\[k_z=\sqrt{k_0^2-k_\perp^2},\,\,\,\,q_z=\sqrt{k_0^2\varepsilon\mu-k_\perp^2}\]
are the respective longitudinal wavevector components in the 
right-handed medium and the left-handed medium.

Boundary conditions (\ref{bound}) provide the following relations between the amplitudes of the three waves:
\begin{eqnarray}
H^{(in0)}e^{ik_zz_l}+H^{(ref1)}e^{-ik_z z_l}=H^{(tr1)}e^{iq_zz_l}~,\nonumber\\
k_zH^{(in0)}e^{ik_zz_l}-k_zH^{(ref1)}e^{-ik_z z_l}={q_z\over\varepsilon}
H^{(tr1)}e^{iq_zz_l}~.
\label{eq8}
\end{eqnarray}

Equations (\ref{eq8}) lead to the expressions for $H^{(ref1)}$ and $H^{(tr1)}$ 
that solve the problem:
\begin{eqnarray}
H^{(tr1)}={2\varepsilon k_z\over{\varepsilon
k_z+q_z}}e^{i(k_z-q_z)z_l}H^{(in0)}~,\nonumber\\
H^{(ref1)}={\varepsilon k_z-q_z\over{\varepsilon k_z+q_z}}e^{2ik_zz_l}H^{(in0)}~.
\label{eq9}
\end{eqnarray}

We note the denominators in Eqs.~(\ref{eq9}). If both the 
incident wave and the transmitted wave are propagating 
ones, then the denominators are nonvanishing at any values 
of $k_\perp$, because ${\rm sgn}\,\varepsilon={\rm sgn}\,q_z$. But if either the incident wave 
or the transmitted wave is a nonpropagating one, for which 
$k_\perp^2>k_0^2,\,k_\perp^2>k_0^2\mu\varepsilon$, then the denominators can vanish:
\begin{equation}
\varepsilon
k_z+q_z\equiv\varepsilon\sqrt{k_0^2-k_\perp^2}+\sqrt{k_0^2\varepsilon\mu-k_\perp^2}=0~.
\label{eq10}
\end{equation}
The signs in front of the roots must be chosen such that both 
the transmitted and the reflected waves decay with increasing 
the distance from the interface, i.e., such that Eq.~(\ref{eq10}) can be 
written as 
\begin{equation}
|\varepsilon|\sqrt{k_\perp^2-k_0^2}-\sqrt{k_\perp^2-k_0^2\varepsilon\mu}=0~.
\label{eq11}
\end{equation}
Equation (\ref{eq11}) is the dispersion equation of {\it surface eigenmodes}, which can exist on the interface of a right-handed 
medium and a left-handed one in the absence of a source field 
$H^{(in0)}$ \cite{Ruppin}.

If the wavevector $k_\perp$ of the source field coincides with the 
solution to dispersion equation (\ref{eq11}), then the amplitudes of 
the transmitted and the reflected waves become infinite, 
because the external field acts as a resonant force exciting 
nondecaying eigenmodes of the medium. Hence, the interface 
itself is a resonant ``amplifier'' of nondecaying waves. Similarly 
to the case of a usual cavity, taking nonzero losses in the 
medium into account restricts the amplitude of the resonant 
surface wave and determines the resonance linewidth. 

Before passing to the plane-parallel lens, we note an 
interesting feature of surface waves: their energy flux is 
directed along the surface and changes sign on the surface. 
In the case of a perfect left-handed medium, the fluxes on the 
right and on the left of the surface have equal absolute values, 
and therefore the total energy flux along the surface is equal to 
zero. 

We now consider a plane wave incident on a plane-parallel 
plate of a left-handed material. Let the second boundary of 
the plane coincide with the plane $z=z_r$. In addition to the 
previously considered waves, two new waves appear now: the 
one reflected from the second boundary and the one passing 
through the plate. Their magnetic fields are denoted by 
\[H^{(ref2)}\exp{(-iq_zz)},\,\,\,H^{(tr2)}\exp{(ik_zz)},\]
respectively. From the boundary conditions on the two 
surfaces, taking all waves into account, we express their 
amplitudes as
\begin{eqnarray}
H^{(tr1)}={1\over 2D(k_\perp)}\left(1+{k_z\varepsilon\over
q_z}\right)e^{-iq_zz_r+ik_zz_l}~,\nonumber\\
H^{(ref1)}=-{i\over 2D(k_\perp)}\left({k_z\varepsilon\over
q_z}-{q_z\over k_z\varepsilon}\right)\sin(q_zd)~,\nonumber\\
H^{(tr2)}={1\over D(k_\perp)}e^{-ik_zd}~,\nonumber\\
 H^{(ref2)}={1\over
2D(k_\perp)}\left(1-{k_z\varepsilon\over q_z}\right)e^{iq_zz_r+ik_zz_l}~,
\label{eq12}
\end{eqnarray}
where $d=z_r-z_l$ is the plate thickness and
\begin{equation}
D(k_\perp)=\cos(q_zd)-{i\over 2}\left({k_z\varepsilon\over q_z}+{q_z\over
k_z\varepsilon}\right)\sin(q_zd)~.
\label{eq13}
\end{equation}

The zeroes of the function $D(k_\perp)$, whenever they exist, 
indicate the existence of nondecaying eigenmodes of the field 
in the plate. It is clear that $D(k_\perp)$ has no zeroes if the 
wavevector $k_z$ is real. This is why we saw no amplification in 
the preceding sections, where we only considered propagating 
waves. 

For an arbitrary sign of $\varepsilon$, the equation $D(k_\perp)=0$ always 
has a solution if $q_z$ is real and $k_z$ is imaginary. These are the 
eigenmodes of a transparent plate placed into a nontransparent medium without dissipation. The nonpropagating field of the source is ``amplified'' near the values of resonant 
wavevectors $k_\perp^{(n)}$ corresponding to the eigenmodes of the 
plate. 

But the most interesting result is obtained when both 
$k_z$ and $q_z$, are imaginary. In this case, the 
$D(k_\perp)$ can be written as \cite{Smith2}
\begin{equation}
D(k_\perp)={1\over 2}e^{-|q_z|d}\left[1-{1\over
2}\left({|k_z|\varepsilon\over|q_z|}+{|q_z|\over|k_z|\varepsilon}\right)\right]+{1\over
2}e^{|q_z|d} \left[1+{1\over
2}\left({|k_z|\varepsilon\over|q_z|}+{|q_z|\over|k_z|\varepsilon}\right)\right]~.
\label{eq14}
\end{equation}
To avoid cumbersome formulas, we analyze Eq.~(\ref{eq14}) for 
$\varepsilon\mu=1$. In this case,
\[|k_z|=|q_z|=\sqrt{k_\perp^2-k_0^2}\]
and the equation $D(k_\perp)=0$ becomes
\begin{equation}
D(k_\perp)\equiv {1\over 2}e^{-\sqrt{k_\perp^2-k_0^2}d}\left[1-{1\over
2}\left(\varepsilon+\varepsilon^{-1}\right)\right]+{1\over
2}e^{\sqrt{k_\perp^2-k_0^2}d}\left[1+{1\over
2}\left(\varepsilon+\varepsilon^{-1}\right)\right]=0~.
\label{eq15}
\end{equation}
It is easy to see that Eq.~(\ref{eq15}) has a solution $k_\perp=k_\perp^{(res)}$ only at negative $\varepsilon$.

The larger the ``amplification'' of a nonpropagating wave, 
the closer $k_\perp$ is to its resonant value $k_\perp^{(res)}$. In other words, the
smaller $D(k_\perp)$, the closer the wave to its resonance and, hence, 
the larger its amplitude at the output of the ``cavity''. As 
$\varepsilon\rightarrow -1$, the function $D(k_\perp)$ decays exponentially fast as $k_\perp$ 
approaches $k_\perp^{(res)}$ from below and is exponentially large above 
the resonance point. Therefore, by convention, we can 
consider the domain $k_0<k_\perp<k_\perp^{(res)}$ to be close to the resonance and the domain $k_\perp>k_\perp^{(res)}$ to be far from the resonance. 

From expressions (\ref{eq12}) and (\ref{eq15}), we obtain the field of the 
nonpropagating Fourier harmonic beyond the plane-parallel 
plate,
\begin{eqnarray}
H^{(tr2)}(z)= e^{\sqrt{k_\perp^2-k_0^2}(d-z)}\left\{{1\over
2}e^{-\sqrt{k_\perp^2-k_0^2}d}\left[1-{1\over
2}\left(\varepsilon+\varepsilon^{-1}\right)\right]+\right.\nonumber\\
\left.{1\over 2}e^{\sqrt{k_\perp^2-k_0^2}d}\left[1+{1\over
2}\left(\varepsilon+\varepsilon^{-1}\right)\right] \right\}^{-1}~.
\label{eq16}
\end{eqnarray}

\begin{figure}[tbh]\centering
\scalebox{0.4}{\includegraphics{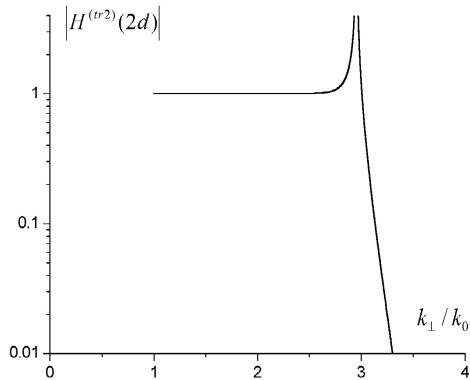}}
\caption{The amplitude of the non-propagating wave in the focal plane of 
a lens made of a left-handed material as a function of the wavevector 
[Eq.(\ref{eq16})]}
\label{Fig8}
\end{figure}

It follows from (\ref{eq16}) that the field of the Fourier harmonic 
in the plane $z=2d$ is close to unity, i.e., to its value in the 
source plane $z=0$, if the wavevector $k_\perp$ is in the ``resonance'' 
domain $k_0<k_\perp<k_\perp^{(res)}$. At the same time, it is exponentially 
small in the domain $k_\perp>k_\perp^{(res)}$
(see Fig.~\ref{Fig8})\footnote{When this paper was in preparation, Ref.~\cite{Rao} appeared where the 
dependence shown in Fig.~\ref{Fig8} was obtained from a numerical solution to the 
Maxwell equations. Because the case of a dissipative medium was 
considered in Ref.~\cite{Rao}, the dependence there had a narrow maximum at 
$k_\perp=k_\perp^{(res)}$ instead of a divergence, as in Fig.~\ref{Fig8}}. We see that in the plane $z=2d$, the amplitudes and phases of the source field are 
the same as in the plane of the source, not only for 
propagating Fourier harmonics, but also for nonpropagating ones. In other words, both propagating and nonpropagating Fourier harmonics participate in the image 
formation. The size of the smallest details resolved in the 
image is given by the value $2\pi/ k_\perp^{(res)}$, which can be much 
smaller than the wavelength $\lambda=2\pi/ k_0$. As $\varepsilon\rightarrow -1$, 
$k_\perp^{(res)}\rightarrow\infty$, and we obtain a ``perfect'' lens. 

In concluding this section, we note the following. In the 
resonance wavevector domain $k_0<k_\perp<k_\perp^{(res)}$, the first term 
in Eq.~(\ref{eq15}), the exponentially small one, is large compared to 
the second term. This is possible only if $\varepsilon$ differs from $-1$ by an 
exponentially small value. Assuming $\varepsilon=-1+\delta$, $|\delta|\ll 1$, we 
obtain from Eq.~(\ref{eq15}) that 
\begin{equation}
k_\perp^{(res)}=k_0\sqrt{1+{\ln^2(2/|\delta|) \over k_0^2d^2}}.
\label{eq17}
\end{equation}

It follows from Eqn.~(\ref{eq17}) that for a given increase in the 
resolving power $k_\perp^{(res)}/k_0$ of the lens, $\delta$ should decrease 
exponentially fast with the growing thickness $d$ of the lens:
\[|\delta|=2\exp\left\{-k_0 d\sqrt{\left({k_\perp^{(res)} \over k_0}\right)^2-1}\right\}.\]

For instance, for $k_\perp^{(res)}/k_0=3$ and $d=\lambda=2\pi/k_0$ (this is 
the case for the dependence in Fig.~\ref{Fig8}), $\delta$ should be smaller than 
$10^{-7}$. Therefore, although a source can formally be imaged 
with any given precision, in practice only a small resolution 
increase is possible for a source placed at a small distance 
($z\simeq\lambda$) from the surface of a thin lens ($d\simeq\lambda$). Such a 
combination of the object, lens, and image placed at a 
distance of the order of the wavelength or smaller should be 
considered not as a lens but rather as a complex source of 
radiation in a layered medium \cite{Veselago3}. 

We make one more remark. For the cavity formed by the 
plate of a left-handed material to amplify the weak ``input 
signal'' (the nonpropagating wave from the source), this cavity 
should accumulate a large energy, which requires a long 
transient process. The larger the resolution that the lens is 
required to produce, the smaller the ``input'' level of the 
corresponding nonpropagating modes, and the larger the 
level of the field in the cavity and the time of the transient 
process. Most probably, ignoring this fact prevented the 
authors of Ref.~\cite{Loshiapo} from observing the effect of a ``perfect'' 
lens when they performed numerical simulation of a {\it time-limited wave packet} propagating through a plane-parallel 
plate of a left-handed material. 

\section{Conclusion}

In this work, we tried to give a review of left-handed media 
and some interesting effects related to them. Certainly, we 
could not mention most of the papers connected with this 
subject (see Fig.~\ref{Fig1}) but we hope to have mentioned the basic 
ones. Although the stream of papers is still huge (in 2003, 
several dozen papers have been published), the peak of the 
excitement has gone. In 1999 -- 2002, almost every new 
statement was soon followed by arguments disproving it, 
but the present discussion is more productive. Now, very few 
researchers doubt the existence of left-handed media. Left-handed media have been realized in various ways by many 
experimentalists, and these works confirm both the unusual 
refraction law at the interface of two media and the focusing 
properties of a plane-parallel plate. The situation is somewhat 
worse with the perfect lens effect, because observation of this 
effect is possible only under very strict requirements to the 
parameters of the left-handed medium. 

In this work, we did not touch upon subjects such as 
nonlinear optical effects at the interface of right-handed and 
left-handed media, one-dimensional periodic waveguide 
media containing left-handed elements, etc. All these ques- 
tions are related to the further development of this direction, 
while our aim was only to acquaint the reader with the basic 
principles of the physics of left-handed media.

This work was supported in part by the INTAS (grant 
No. 03-55-1921).

\end{document}